\documentclass[ reprint,  amsmath,amssymb,  aps, ]{revtex4-1}%
\usepackage{graphicx}
\usepackage{dcolumn}
\usepackage{bm}
\usepackage{amsmath}
\usepackage{amsfonts}
\usepackage{amssymb}
\usepackage{color}

\providecommand{\U}[1]{\protect\rule{.1in}{.1in}}

\begin{document}

\title{
High-Field Ultrasonic Study of Quadrupole Ordering and Crystal Symmetry Breaking in CeRhIn$_5$
}

\author{
R. Kurihara$^1$, 
A. Miyake$^1$,
M. Tokunaga$^1$,
Y. Hirose$^2$,
and
R. Settai$^2$
}

\affiliation{
$^1$Institute for Solid State Physics, The University of Tokyo,  Kashiwa, Chiba 277-8581, Japan \\
$^2$Department of Physics, Niigata University, Niigata 950-2181, Japan
}

\begin{abstract}
We performed an ultrasonic measurement for the heavy-fermion compound CeRhIn$_5$ to investigate the origin of the field-induced anisotropic phase in high magnetic fields.
The transverse elastic constant $C_\mathrm{T} =  (C_{11} - C_{12})/2$ and the ultrasonic attenuation coefficient $\alpha_\mathrm{T}$ show clear anomaly at $B^\star = 28.5$ T, which was discussed as the electronic nematic transition point.
In addition, $C_\mathrm{T}$ exhibits acoustic de Haas-van Alphen oscillation below 28.5 T.
These elastic anomalies around $B^\star$ indicate an electric quadrupole ordering of $O_{x^2-y^2}$ accompanied by $B_\mathrm{1g}$ crystal symmetry breaking and Fermi surface reconstruction due to the quadrupole-strain coupling, which results from itinerant $4f$ electrons and the $p$-$f$ hybridized state.
\end{abstract}

\maketitle
%%%%%%%%%%%%%%%%Affil%%%%%%%%%%%%%%%%%%%%%%%%%%%%%

%%%%%%%%%%%%%%%%ABST%%%%%%%%%%%%%%%%%%%%%%%%%%%%

%%%%%%%%%%%%%%%%%%%%%    Introduction     %%%%%%%%%%%%%%%%%%%%%%%%%%%%%%%%%%
\section{
\label{sect_intro}
Introduction
}

Symmetry breaking is an important concept in the description of phase transitions.
Time-reversal, spatial-inversion, U(1) gauge symmetry breaking appear in ferromagnetic, ferroelectric, and superconducting transitions.
Recently, the in-plane anisotropic state, which is characterized by the lacking of the $\pm \pi/2$-rotational operation $C_4^{\pm1}$ from the high-symmetry space- and point-group, is called as electronic nematic (EN) state.
The EN state has been observed in a number of strongly correlated electron systems
\cite{Borzi_Science315, Ando_PRL88, Chuang_Science327, Chu_Science329, Goto_JPSJ80, Yoshizawa_JPSJ81, Kurihara_JPSJ86}.
Furthermore, the nematic contribution to a structural phase transition, superconductivity, and other exotic phenomena have been studied as well
\cite{Yanagi_JPSJ79, Kontani_PRB84, Kontani_PRL113}.
Magnetic field-induced EN transition in the heavy-fermion compound of CeRhIn$_5$ has also been observed as an in-plane anisotropic magnetoresistance
\cite{Moll_NatureComm7663, Ronning_Nature313}.

The structural, magnetic, and electronic properties of CeRhIn$_5$ have been investigated by many methods.
The CeRhIn$_5$ compound, with a HoCoGa$_5$ type crystal structure belonging to the $P4/mmm$ ($D_{4h}^1$) space group 
\cite{Hegger_PRL84},
exhibits an antiferromagnetic (AFM) transition at $T_\mathrm{N} = 3.8$ K with a helical magnetic structure
\cite{Hegger_PRL84, Bao_PRB62}.
In the AFM phase, in-plane magnetization measurements have shown a metamagnetic transition at $B_\mathrm{m} =2$ T and the disappearance of AFM at $B_0 = 50$ T, while the magnetization for $[001]$ shows a monotonic increase up to 52 T 
\cite{Takeuchi_JPSJ70}.
A de Haas-van Alphen (dHvA) measurement demonstrated that CeRhIn$_5$ had localized $4f$ electron compared with the non-$4f$ reference compound LaRhIn$_5$
\cite{Shishido_JPSJ71}.

The application of hydrostatic pressure suppresses the AFM order at $2.1$ GPa.
Change of the Fermi surface (FS) occurs at $P_\mathrm{c} = 2.35 $ GPa as indicated by the change in the dHvA frequency and the effective mass enhancement
\cite{Shishido_JPSJ74}.
Superconductivity appears at $1.5$ GPa and is the most stable with the transition temperature of $T_\mathrm{c} = 2.2$ K at the quantum critical point of $2.4$ GPa
\cite{Hegger_PRL84, Knebel_PRB74}.
These pressure-induced properties can be attributed to the change from the localized to itinerant $4f$-electron character
\cite{Doniach_PhysicaB91}.

While CeRhIn$_5$ has been treated as a localized system at ambient pressure, the $4f$ electrons of the related compounds of CeCoIn$_5$ and CeIrIn$_5$ exhibit itinerant property with a huge effective mass and superconductivity at low temperatures
\cite{Shishido_JPSJ74, Settai_JPCM13, Petrovich_EPL53, Petrovich_JPCM13}.
The difference between localized and itinerant properties in these Ce-115 compounds has been described in terms of the out-of-plane orbital anisotropy $\alpha^2$ in the ground-state wavefunction under crystalline electric field (CEF) as
$ \left|\Gamma_7^{\mathrm{G} \pm} \right> = \alpha\left| J_z = \pm 5/2 \right>+\sqrt{1-\alpha^2}\left| J_z = \mp 3/2 \right>$,
where
$\left| J_z = \pm 5/2 \right>$ has a donut shape and $\left| J_z = \pm 3/2 \right>$ has a yo-yo shape
\cite{Willers_PNAS112, Chen_PRB97, Willers_PRB81}.
The out-of-plane orbital contribution can be tuned by materials as CeRhIn$_5$ ($\alpha^2 = 0.38$), CeIrIn$_5$ ($0.25$), and CeCoIn$_5$ ($\alpha^2 = 0.13$).
Thus the latter has a stronger three-dimensional (3D) character.
Here, the smaller $\alpha^2$ represents the stronger three dimensional (3D) character.
In addition to the $\alpha^2$ scaling, the $4f$ itinerancy due to hybridization between the Ce-4$f$ and the out-of-plane In-5$p$ electrons has been theoretically discussed
\cite{Haule_PRB81}.
Thus, in the Ce-115 system, the 3D CEF ground state and the $p$-$f$ hybridization studied for several Ce-based compounds
\cite{Kasuya_JPhysC18_2697, Kasuya_JPhysC18_2709, Kasuya_JPhysC18_2721}
are important to understand the $4f$ delocalization due to the hydrostatic pressure and the substitution of Co, Rh, and Ir.

In the AFM phase of CeRhIn$_5$ in fields greater than $B^\star \sim 30$ T, the in-plane anisotropic state accompanied by anisotropic electronic properties have been observed.
The magnetoresistance measurements under $B // [001]$ have revealed anisotropies between the resistivity of the $[110]$ and $[1\bar{1}0]$ directions, as well as the $[100]$ and $[010]$ directions, which are equivalent under the $C_4^{\pm1}$ operations of the tetragonal crystal
\cite{Moll_NatureComm7663, Ronning_Nature313}.
The magnetostriction measurements have demonstrated the anomaly due to the in-plane anisotropy at $B^\star$
\cite{Rosa_PRL122, Jiao_PRB99}.
The hybridization of Ce-$4f$ and in-plane In-$5p$ electrons, which was enhanced by the increase in $\alpha^2$ parameter due to the CEF wave function mixing between the ground and first excited states, was also suggested as the origin of the in-plane anisotropic state. 
At $B^\star$, the dHvA effect has revealed FS reconstruction in terms of volume change
\cite{Jiao_PRB99}.
In addition, high-field specific heat measurements have revealed mass enhancement 
\cite{Jiao_PNAS112}.
These FS reconstruction and mass enhancement were explained in terms of the itinerancy of $4f$ electrons.  
Consequently, the field-induced EN property and FS reconstruction indicate the importance of the in-plane $p$-$f$ hybridization and the delocalization of the $4f$ electrons.

It is crucially important to unambiguously identify the order parameter and the electronic state of the proposed field-induced EN phase in CeRhIn$_5$.
However, the active representation, which describes the symmetry breaking of the field-induced EN phase in CeRhIn$_5$, remains ambiguous because there are two irreducible representations of $B_\mathrm{1g}$ and $B_\mathrm{2g}$ describing the lack of $C_4^{\pm1}$ operation. 
To characterize the symmetry breaking, we focus on the ultrasonic properties.
It is a powerful tool to determine the active representation of a phase transition related to the crystal symmetry breaking because an ultrasonic wave can induce and identify both the symmetry strain $\varepsilon_{x^2-y^2}$ with $B_\mathrm{1g}$ and $\varepsilon_{xy}$ with $B_\mathrm{2g}$ as listed in Tab. \ref{table1}.
In addition, we can propose the electric quadrupole as an order parameter of the crystal symmetry breaking in terms of the quadrupole-strain interaction, which is based on the selection rule of group theory.
An electronic state inducing the quadrupole ordering can also be discussed to calculate the expectation value of an electric quadrupole.
These ultrasonic properties have shown the importance of the quadrupole, which originates from the orbital degree of freedom of the electron on the FS, for example, in the structural phase transition and in-plane anisotropy in iron pnictide superconductors
\cite{Goto_JPSJ80, Yoshizawa_JPSJ81, Kurihara_JPSJ86}
as well as in the lattice instability of URu$_2$Si$_2$
\cite{Yanagisawa_PRB97}.
The ultrasonic properties have also been used to investigate field-induced quadrupole ordering in $4f$-electron compounds
\cite{Nakamura_PhysicaB_199and200, Nakamura_JPSJ64, Ishii_PRB97}.

This paper is organized as follows.
In Sect. \ref{sect_exp}, the experimental procedures of sample preparation, ultrasonic measurements, and pulsed magnetic fields are described.
In Sect. \ref{sect_3}, we present the results of the ultrasonic experiments, which indicate $B_\mathrm{1g}$ crystal symmetry breaking and FS reconstruction due to the electric quadrupole ordering of $O_{x^2-y^2}$ at the proposed EN phase.
The field dependence of the elastic constants and that of the acoustic de Haas-van Alphen oscillations are discussed.
In Sect. \ref{sect_Quad}, the possible electronic states originating from the electric quadrupole degree of freedom are discussed.
We conclude our results in Sect. \ref{conclusion}.

%%%%%%%%%%%%%%%%%%%%%    Experiment     %%%%%%%%%%%%%%%%%%%%%%%%%%%%%%%%%%
\section{
\label{sect_exp}
Experiment
}

Single crystals of CeRhIn$_5$ were grown by the flux method.
To investigate the active representation of the proposed EN state, two samples were prepared: one with (100) and ($\bar{1}$00) faces, and another with (110) and ($\bar{1}$$\bar{1}$0) faces.
The ultrasonic pulse-echo method with a numerical vector-type phase detection technique was used for the ultrasonic velocity $v$ and for the ultrasonic attenuation coefficient $\alpha_\mathrm{T}$
\cite{Fujita_JPSJ80}.
Piezoelectric transducers using LiNbO$_3$ plates with a 36$^\circ$ Y-cut and an X-cut were employed to generate longitudinal ultrasonic waves with the fundamental frequency of approximately $f = 30$ MHz and the transverse waves with 16 MHz, respectively. 
Higher-harmonic frequencies of 68 MHz and 112 MHz were also employed for the acoustic de Haas-van Alphen oscillation and the $\alpha_\mathrm{T}$ measurements, respectively.
The elastic constant $C = \rho v^2$ was calculated from the ultrasonic velocity $v$ and the mass density of $\rho = 8.316$ g/cm$^3$.
The direction of ultrasonic propagation, $\boldsymbol{q}$, and the direction of polarization, $\boldsymbol{\xi}$, for the elastic constant $C_{ij}$ are indicated in all figures in the paper.
For high-field measurements up to 56 T, a nondestructive pulse magnet with a time duration of 36 ms installed at The Institute for Solid State Physics, The University of Tokyo was used. 

%%%%%%%%%%%%%%%%%%%%%%%%Fig1%%%%%%%%%
\begin{figure}[t]
\begin{center}
\includegraphics[clip, width=0.5\textwidth]{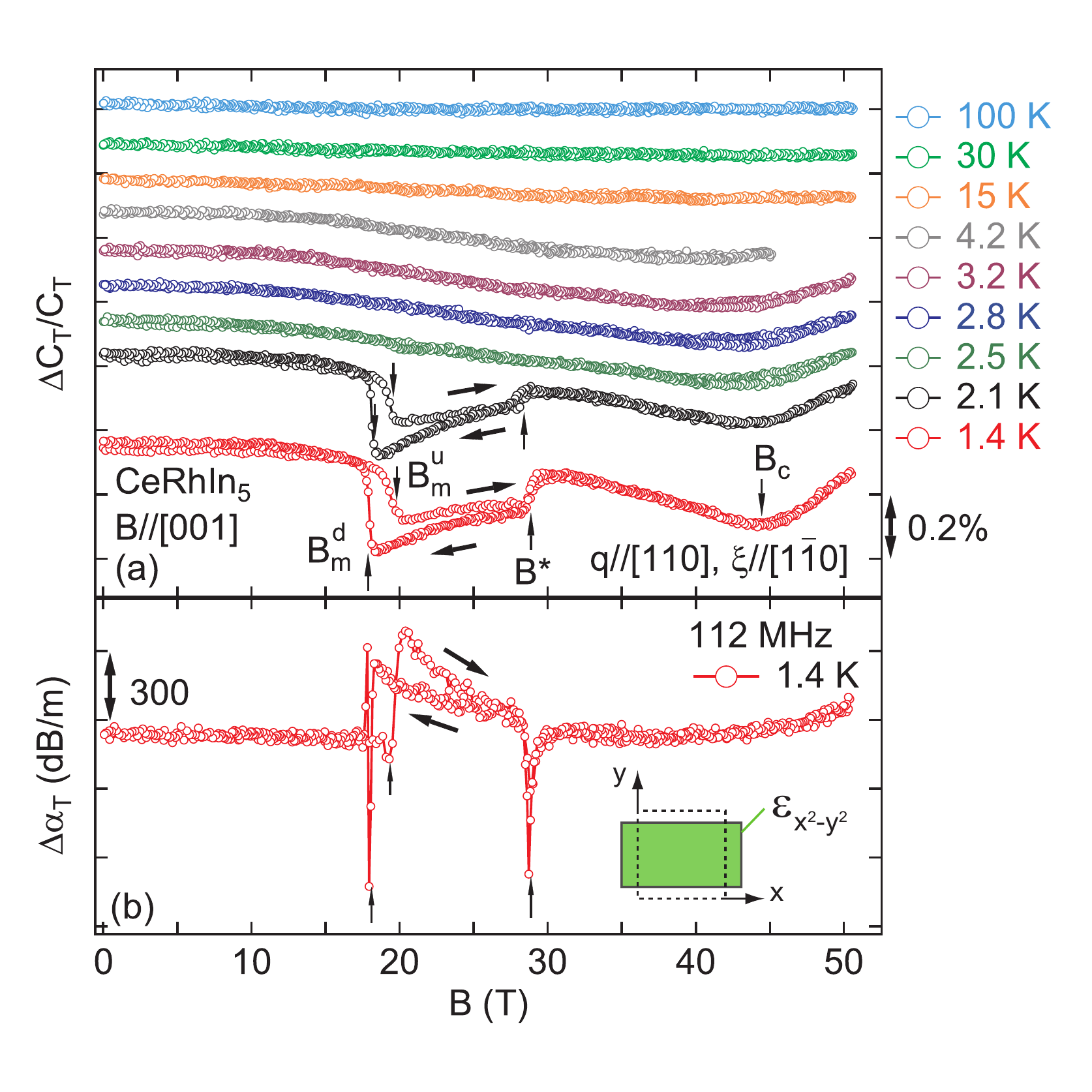}
\end{center}
\caption{
(Color online)
Transverse elastic constant $C_\mathrm{T} = \left(C_{11} - C_{12} \right)/2$ describing the $B_\mathrm{1g}$ symmetry breaking of $D_{4h}$ in CeRhIn$_5$.
The absolute value of $C_\mathrm{T}$ at 4.2 K is estimated to be $4.7 \times 10^{10}$ J/m$^3$.
(a) Magnetic field dependence of the elastic constant $\mathit{\Delta}C_\mathrm{T}/C_\mathrm{T}$ at several temperatures for $B // [001]$.
The vertical arrows indicate the metamagnetic transition field $B_\mathrm{m}^\mathrm{u(d)}$ for field up(down)-sweep and the EN transition field $B^\star$.
The right- and left-arrows show hysteresis directions.
(b) Magnetic field dependence of the ultrasonic attenuation coefficient $\mathit{\Delta} \alpha_\mathrm{T}$.
In the inset of panel (b), the dashed square and the green rectangle indicate the tetragonal unit-cell and deformed cell due to the strain $\varepsilon_{x^2-y^2}$, respectively.
}
\label{Fig1}
\end{figure}
%%%%%%%%%%%%%%%%%%%%%%%%%%%%%%%

%%%%%%%%%%%%%%%%%%%%%   Results    %%%%%%%%%%%%%%%%%%%%%%%%%%%%%%%%%%
\section{
\label{sect_3}
Results and discussion
}

\subsection{
\label{subsect_CT}
Transverse elastic constant $(C_{11}-C_{12})/2$
}
To identify the active representation of the in-plane symmetry breaking accompanied by the proposed EN transition, the five elastic constants of CeRhIn$_5$ were measured under pulsed magnetic fields applied along the [001] direction.
In this section, we discuss the transverse elastic constant $C_\mathrm{T} =  (C_{11} - C_{12})/2$ and the ultrasonic attenuation coefficient $\alpha_\mathrm{T}$ related to the symmetry breaking of the irreducible representation $B_\mathrm{1g}$
\cite{Luthi Phys. Ac., Inui_group}.

Figure \ref{Fig1}(a) shows the magnetic field dependence of
$\mathit{\Delta}C_\mathrm{T}/C_\mathrm{T} = \left[ C_\mathrm{T} \left( B \right) - C_\mathrm{T} \left( B = 0 \right) \right] /C_\mathrm{T} \left( B = 0 \right)$
at several temperatures.
We observed several anomalies in the $C_\mathrm{T} \left( B \right)$ curves below 2.1 K.
At 1.4 K, $\mathit{\Delta}C_\mathrm{T}/C_\mathrm{T}$ exhibits an elastic softening of $2.3 \times 10^{-3}$ with the increase in the fields, as it approaches $B_\mathrm{m}^\mathrm{u} = 19.5$ T, where the metamagnetic transition takes place due to the misalignment of the magnetic fields from the $[001]$ direction
\cite{Shishido_JPSJ71, Jiao_PNAS112}.
Considering $B_\mathrm{m} = 2$ T for the in-plane fields, we estimated a tilting angle $\theta$, measured from the [001] direction to a given in-plane direction, to be $6.2^\circ$ by the function
$B_\mathrm{m} \left( \theta \right) = B_\mathrm{m} \left(\theta = 90^\circ \right)/\cos \left( 90^\circ - \theta \right)$.
Above $B_\mathrm{m}^\mathrm{u}$, $C_\mathrm{T}$ shows a hardening and rapid increasing at $B^\star  = 28.5$ T.
The anomaly at $B^\star$ can be attributed to the EN transition compared with those in previous reports
\cite{Moll_NatureComm7663, Ronning_Nature313, Rosa_PRL122, Jiao_PNAS112}.
With the further application of the fields, $C_\mathrm{T}$ has a minimum at $B_\mathrm{c} = 44.6$ T.
With the decrease in the fields, hysteresis behavior appears in $C_\mathrm{T}$ below $B^\star$.
$C_\mathrm{T}$ also shows a rapid increase at the metamagnetic transition field of $B_\mathrm{m}^\mathrm{d} = 18.0$ T.
At 2.1 K, the $C_\mathrm{T} \left( B \right)$ curve shows almost the same profile to that at 1.4 K.

The metamagnetic and the EN phase transitions cannot be resolved above 2.5 K.
In the AFM phase at 2.5 K, 2.8 K, and 3.2 K, $C_\mathrm{T}$ shows a monotonic softening up to $B_\mathrm{c}$ without any anomalies observed at lower temperatures as shown in Fig. \ref{Fig1}(a).
In paramagnetic (PM) state above $T_\mathrm{N} = 3.8$ K, $C_\mathrm{T}$ shows monotonic softening with the increase in the fields.
Therefore, it is expected that $B_\mathrm{c}$ is located above 56 T in the PM phase above 15 K. 

The anomalies at $B_\mathrm{m}^\mathrm{u}$, $B_\mathrm{m}^\mathrm{d}$, and $B^\star$ also appear in the ultrasonic attenuation coefficient of $C_\mathrm{T}$ mode, $\alpha_\mathrm{T}$.
Figure \ref{Fig1}(b) shows the magnetic field dependence of
$\mathit{\Delta} \alpha_\mathrm{T} = \alpha_\mathrm{T} \left( B \right) - \alpha_\mathrm{T} \left( B = 0 \right) $ at 1.4 K.   
It can be seen that $\mathit{\Delta} \alpha_\mathrm{T}$ with 112 MHz shows step-like change across $B_\mathrm{m}^\mathrm{u}$ and $B_\mathrm{m}^\mathrm{d}$, and a sharp dip at $B^\star$.

%%%%%%%%%%%%%%%%%%%%%%%%Fig2%%%%%%%%%
\begin{figure}[t]
\begin{center}
\includegraphics[clip, width=0.5\textwidth]{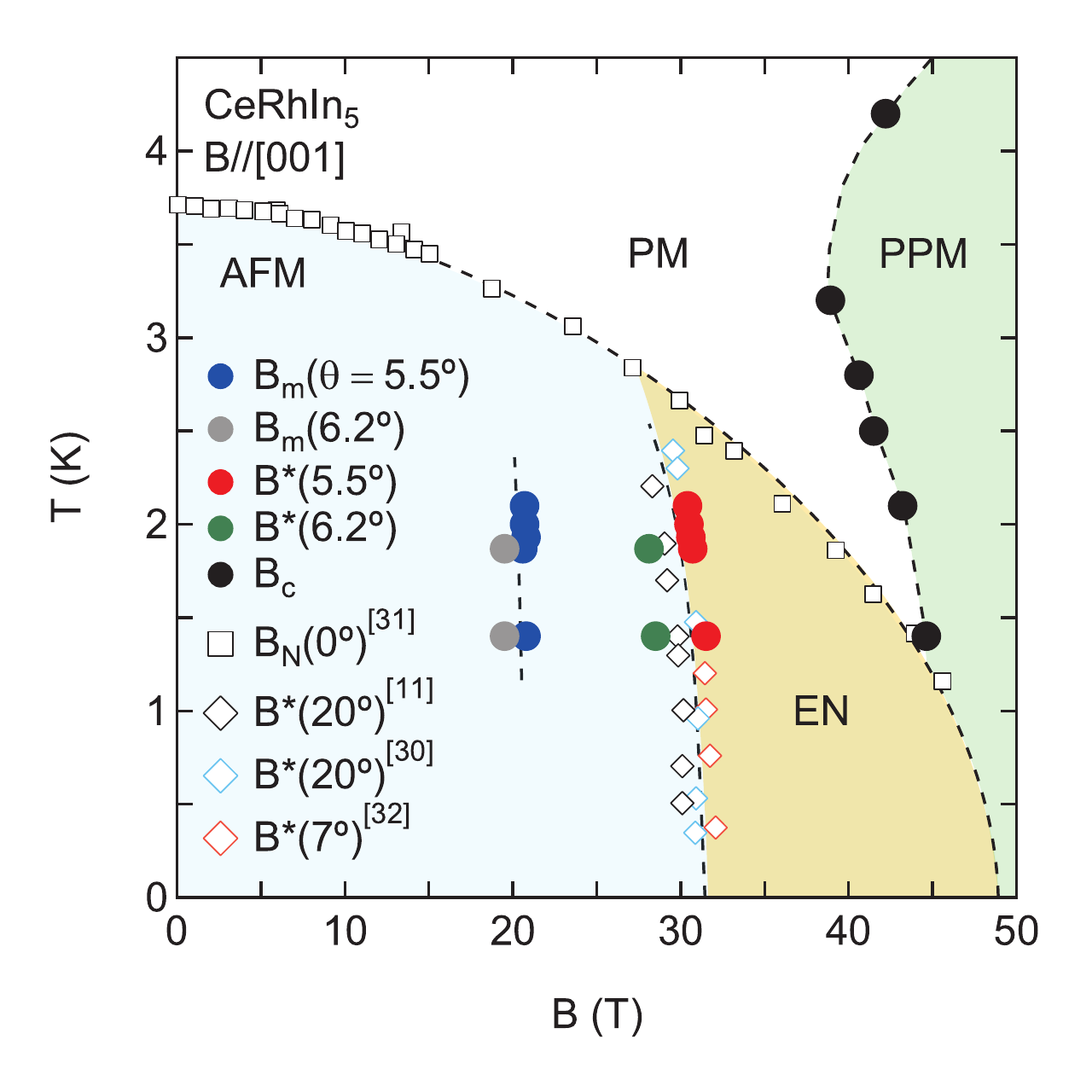}
\end{center}
\caption{
(Color online)
Temperature-field phase diagram of CeRhIn$_5$ decided by the transverse elastic constant $(C_{11}-C_{12})/2$ for $B / [001]$.
The EN transition field $B^\star$ at the field tilting angles $\theta = 5.5^\circ$ and $6.2^\circ$ are shown by the filled red and green circles, respectively.
The metamagnetic transition field $B_\mathrm{m}$ at $5.5^\circ$ and $6.2^\circ$ are indicated by the filled blue and gray circles, respectively.
The minimum field $B_\mathrm{c}$ is shown by the filled black circles.
The open black square indicates $B_\mathrm{N}$ as obtained in Ref. \cite{Jiao_PRB99}.
The open black, blue, and red rhombus show $B^\star$ in Refs. \cite{Ronning_Nature313, Rosa_PRL122, Jiao_PNAS112}, respectively.
}
\label{Fig2}
\end{figure}
%%%%%%%%%%%%%%%%%%%%%%%%%%%%%%%

We summarized the elastic anomalies in the transverse elastic constant $(C_{11}-C_{12})/2$ at the magnetic phase diagram in Fig. \ref{Fig2}.
The EN transition field $B^\star$ at $\theta = 6.2^\circ$ shown in Fig. \ref{Fig1} can be consistent with the previous results by the magnetoresistance in Refs.
\cite{Ronning_Nature313, Rosa_PRL122, Jiao_PNAS112}.
The anomalies of $B^\star$ at $\theta = 5.5^\circ$, which will be explained in the following Sect. \ref{subsect_AdHvA}, are also consistent with previous reports. 
On the other hand, the elastic minimum at $B_\mathrm{c}$ is quite different from the anomalies of the EN state, the metamagnetic transition, and the AFM boundary.
This origin will be discussed in the following Sect. \ref{subsect_AllC}. 

Our ultrasonic measurements of $C_\mathrm{T}$ and $\alpha_\mathrm{T}$ with the $B_\mathrm{1g}$ irreducible representation of $D_{4h}$ exhibit the elastic anomaly at $B^\star$.
This result suggests that the electronic degree of freedom with $B_\mathrm{1g}$, which describes the field-induced EN transition, couples to the strain $\varepsilon_{x^2-y^2} = \varepsilon_{xx} - \varepsilon_{yy}$ with $B_\mathrm{1g}$ as indicated in the inset of Fig. \ref{Fig1}(b) induced by the ultrasonic waves.
This electronic degree of freedom can be the electric quadrupole $O_{x^2-y^2} = \left( x^2-y^2 \right)/r^2$ shown in Tab. \ref{table1}, because the basis of the $B_\mathrm{1g}$ is described by the form of $x^2-y^2$
\cite{Inui_group}.
This coupling is described by the quadrupole-strain interaction given by
\cite{Luthi Phys. Ac.}
%HQS 
\begin{align}
\label{HQS}
H_\mathrm{QS} = -g_{x^2-y^2}O_{x^2-y^2}\varepsilon_{x^2-y^2}.
\end{align}
Here, $g_{x^2-y^2}$ is a coupling constant.
An elastic constant and an ultrasonic attenuation coefficient are related to the susceptibility of an electric quadrupole
\cite{Luthi Phys. Ac.}.
Consequently, the anomalies of $C_\mathrm{T}$ and $\alpha_\mathrm{T}$ at $B^\star$ suggest that the proposed field-induced EN transition in CeRhIn$_5$ can be regarded as the ferro-type electric quadrupole ordering of $O_{x^2-y^2}$ accompanying the crystal symmetry breaking, given by the strain $\varepsilon_{x^2-y^2}$ with the $B_\mathrm{1g}$ active representation.

%%%%%%%%%%%%%%%%%%%%%%%%Fig3%%%%%%%%%
\begin{figure}[t]
\begin{center}
\includegraphics[clip, width=0.5\textwidth]{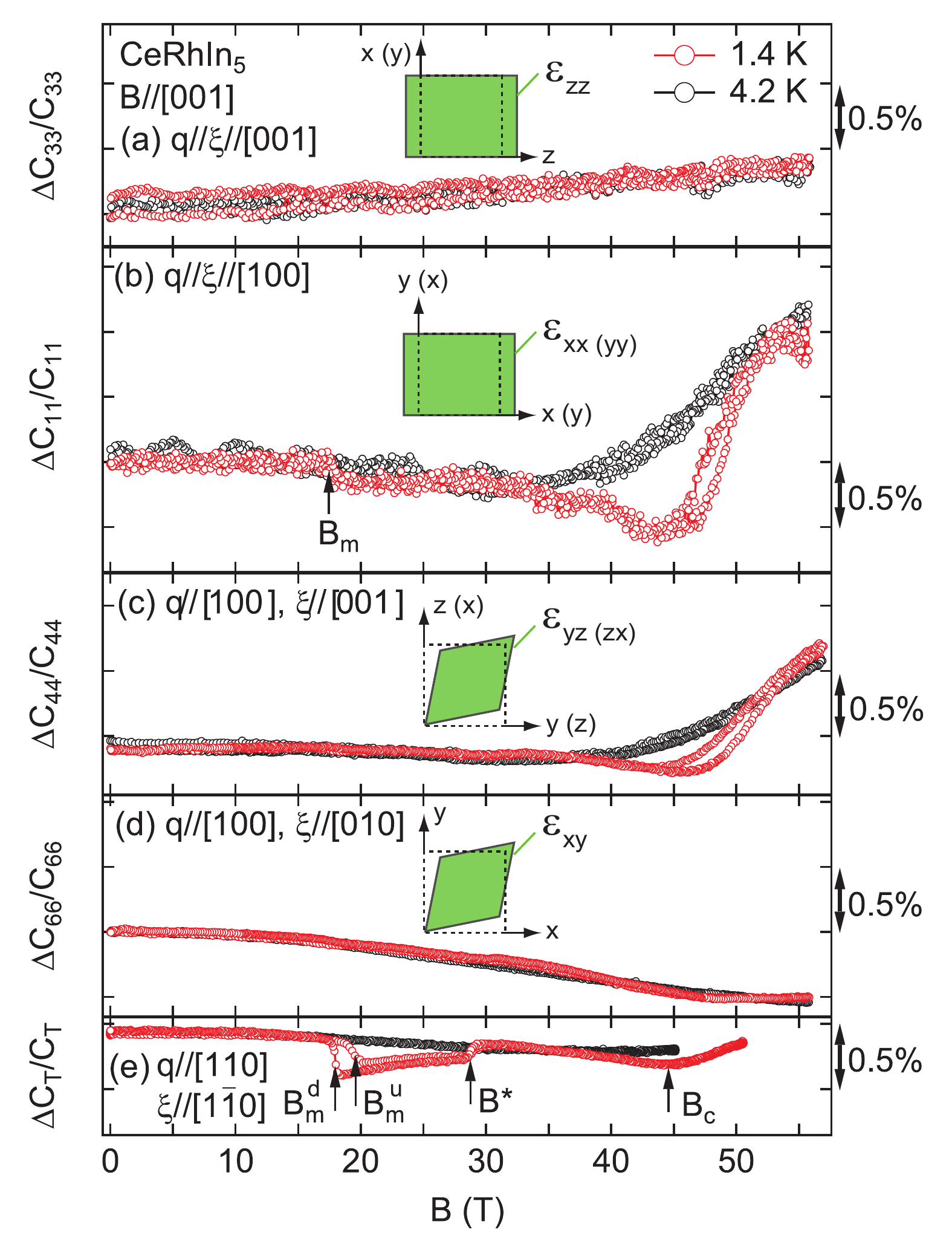}
\end{center}
\caption{
(Color online)
Magnetic field dependence of the relative variation in the elastic constants $\mathit{\Delta}C_{ij}/C_{ij}$ of CeRhIn$_5$ at 1.4 K and 4.2 K for $B // [001]$.
Longitudinal elastic constants
(a) $C_{ij} = C_{33}$ with $6.5 \times 10^{10}$ J/m$^3$ at 4.2 K and 0 T
and (b) $C_{11}$ with $10 \times 10^{10}$ J/m$^3$.
Transverse elastic constants
(c) $C_{44}$  with $3.6 \times 10^{10}$ J/m$^3$,
(d) $C_{66}$ with $3.9 \times 10^{10}$ J/m$^3$, 
and (e) $C_\mathrm{T} =  (C_{11} - C_{12})/2$ with $4.7 \times 10^{10}$ J/m$^3$.
The strain $\varepsilon$ due to the ultrasonic waves for $C_{ij}$ is schematically drawn in the inset.
}
\label{Fig3}
\end{figure}
%%%%%%%%%%%%%%%%%%%%%%%%%%%%%%%

%%%%%%%%%%%%%%%%Table%%%%%%%%%%%%%%%%%
\begin{table*}[htbp]
\caption{
Symmetry strains, electric quadrupoles, and elastic constants corresponding to the irreducible representations (IR) in $D_{4h}$.
In the columns of $B^\star$ and $B_\mathrm{c}$, $\bigcirc$ and $-$ signs indicate whether or not the respective elastic constants show anomaly, respectively. 
}
\begin{ruledtabular}
\label{table1}
\begin{tabular}{cccccc}
\textrm{IR}
& \textrm{Symmetry strain}
	& \textrm{Electric quadrupole}
		& \textrm{Elastic constant}
			& $B^\star$ \textrm{(EN)}
				& $B_\mathrm{c}$
\\
\hline
$A_\mathrm{1g}$
	& $\varepsilon_\mathrm{B} = \varepsilon_{xx} + \varepsilon_{yy} + \varepsilon_{zz}$
		& 
			& $ C_\mathrm{B} = \left( 2C_{11} + 2C_{12} + 4C_{13} + C_{33} \right)/9 $
				& $-$
					& $-$
\\

	& $\varepsilon_u = (2\varepsilon_{zz} - \varepsilon_{xx} - \varepsilon_{yy})/\sqrt{3}$
		& $O_{3z^2-r^2} = \left( 3z^2-r^2 \right)/r^2$
			& $C_u = \left( C_{11} + C_{12} - 4C_{13} + 2C_{33} \right)/6$
				& $-$
					& $-$
\\
$B_\mathrm{1g}$
	& $\varepsilon_{x^2-y^2} = \varepsilon_{xx} - \varepsilon_{yy}$
		& $O_{x^2-y^2} = \left( x^2 -y^2 \right)/r^2$
			& $C_\mathrm{T} = \left( C_{11} - C_{12} \right)/2$
				& $\bigcirc$
					& $\bigcirc$
\\
$B_\mathrm{2g}$
	& $\varepsilon_{xy}$
		& $O_{xy} =  xy/r^2$
			& $C_{66}$
				& $-$
					& $\bigcirc$
\\
$E_\mathrm{g}$
	& $\varepsilon_{yz}$
		& $O_{yz} = yz/r^2$
			& $C_{44}$
				& $-$
					& $\bigcirc$
\\

	& $\varepsilon_{zx}$
		& $O_{zx} = zx/r^2$
			&$C_{44}$ 
				& $-$
					& $\bigcirc$
\end{tabular}
\end{ruledtabular}
\end{table*}
%%%%%%%%%%%%%%%%%%%%%%%%%%%%%%%%%%%%

%%%%%%elastic constants

\subsection{
\label{subsect_AllC}
Elastic constants $C_{33}$, $C_{11}$, $C_{44}$ and $C_{66}$ 
}

For further discussion, we measured the other elastic constants of CeRhIn$_5$ with the tetragonal crystal structure.
Figure \ref{Fig3} shows the magnetic field dependence of the relative elastic constants $\mathit{\Delta}C_{ij}/C_{ij}$ at 1.4 K and 4.2 K.
The significant experimental result to understand the symmetry breaking accompanied by the EN transition only appears in $C_\mathrm{T}$ with $B_{1g}$ shown in Fig. \ref{Fig3}(d).
The other elastic constants do not show any anomaly at $B^\star$.
As  shown in Fig. \ref{Fig3}(a), the longitudinal elastic constant $C_{33}$ at 1.4 K exhibits a monotonic hardening up to 56 T.
In contrast, another longitudinal elastic constant $C_{11}$ at 1.4 K shown in Fig. \ref{Fig3}(b) shows softening on approaching $B_\mathrm{c}$ and a subsequent hardening for further high fields.
A small anomaly also appears at $B_\mathrm{m}$, which is only seen in $C_\mathrm{T}$ and $C_{11}$.
As shown in Fig. \ref{Fig3}(c), the transverse elastic constant $C_{44}$ with $E_\mathrm{g}$ at 1.4 K exhibits similar field dependence to $C_{11}$ except for no anomaly at $B_\mathrm{m}$.
The transverse elastic constant $C_{66}$ with $B_\mathrm{2g}$ at 1.4 K shown in Fig. \ref{Fig3}(d) shows an inflection point around 45 T and a minimum point at 48 T.
Comparing to the other elastic constants, the inflection point corresponds to $B_\mathrm{c}$.

To clarify the symmetry breaking character at the EN phase, we discuss the contribution of the strains in Tab. \ref{table1}.
Our experimental results indicate that the $B_\mathrm{1g}$ can be the active representation of the EN phase.
The strain $\varepsilon_{zz}$ induced by the longitudinal ultrasonic waves for $C_{33}$ is reduced to the bulk strain $\varepsilon_\mathrm{B}$ and the tetragonal strain $\varepsilon_u$ as
$\varepsilon_{zz} = \varepsilon_\mathrm{B}/3 + \varepsilon_u/\sqrt{3}$.
Therefore, both $\varepsilon_\mathrm{B}$ and $\varepsilon_u$ have no contribution to the quadrupole-strain interaction in the field-induced EN phase because of the absence of the anomaly at $B^\star$ in $C_{33}$. 
The longitudinal ultrasonic waves for $C_{11}$ induce the strain $\varepsilon_{xx}$, which is reduced as
$\varepsilon_{xx} = \varepsilon_\mathrm{B}/3 - \varepsilon_u/2\sqrt{3} + \varepsilon_{x^2-y^2}/2$.
Therefore, $C_{11}$ inducing the strain $\varepsilon_{x^2-y^2}$ in part should show the anomaly at $B^\star$.
However, as seen in Fig. \ref{Fig3}(b), the anomaly at $B^\star$ in $C_{11}$ is unclear due to the experimental noise level, which is  comparable to the relative change of $C_\mathrm{T}$ at $B^\star$.
The elastic measurements of $C_{44}$ and $C_{66}$ also indicate no contribution of the strains $\varepsilon_{yz}$ and $\varepsilon_{zx}$ with $E_\mathrm{g}$ and $\varepsilon_{xy}$ with $B_\mathrm{2g}$ to the EN phase.

As discussed above, only the strain $\varepsilon_{x^2-y^2}$ contributes to the anomaly at $B^\star$ as summarized in Tab. \ref{table1}.
Therefore, the field-induced EN transition in CeRhIn$_5$ results from the ferro-type ordering of the electric quadrupole $O_{x^2-y^2}$ with the $B_\mathrm{1g}$ irreducible representation of $D_{4h}$.
In addition, $B_\mathrm{1g}$ crystal symmetry breaking due to the quadrupole-strain interaction given in Eq. (\ref{HQS}) can also be induced.
These electric quadrupole ordering and crystal symmetry breaking are consistent with the in-plane anisotropy of the resistivity of the $[100]$ and $[010]$ directions
\cite{Moll_NatureComm7663, Ronning_Nature313},
the magnetostriction along the [100] direction
\cite{Rosa_PRL122, Jiao_PRB99},
and the absence of anomaly in the magnetization at $B^\star$
\cite{Takeuchi_JPSJ70}.

While the symmetry breaking character for the EN phase is identified, that of the metamagnetic transition remains ambiguous.
In addition to $C_\mathrm{T}$, the anomaly at $B_\mathrm{m}$ appears in $C_{11}$.
This anomaly can be caused due to the misalignment of the magnetic fields from the $[001]$ direction.
In contrast to $C_{11}$ and $C_\mathrm{T}$, the anomaly due to the metamagnetic transition is hardly visible in the elastic constants $C_{33}$, $C_{44}$, and $C_{66}$ shown in Fig. \ref{Fig3}.
At the moment, it is not clear whether this experimental result suggests that the strains $\varepsilon_\mathrm{B}$, $\varepsilon_u$, $\varepsilon_{yz}$, $\varepsilon_{zx}$, and $\varepsilon_{xy}$ are not active for the metamagnetic transition or $B_\mathrm{m}$ becomes larger than $56 $T owing to the field misalignment smaller than $2.0^\circ$.
To understand the symmetry breaking of the metamagnetic transition, we need to measure the field angle dependence of the elastic constants.

Anomalies at $B_\mathrm{c}$ appear in the elastic constants $C_{11}$, $C_{44}$, $C_{66}$, and $C_\mathrm{T}$.
This fact indicates that 3D nature of an electronic state can be the origin of the anomaly because all of the symmetry breaking strains $\varepsilon_{x^2-y^2}$ with $B_{1g}$, $\varepsilon_{yz}$ and $\varepsilon_{zx}$ with $E_{g}$, and $\varepsilon_{xy}$ with $B_{2g}$ exhibit the anomaly at $B_\mathrm{c}$.
In some heavy fermion systems in high magnetic fields, a polarized paramagnetic (PPM) state has been studied.
In the revealed $B$-$T$ phase diagram, the PPM phase boundary shifts to higher fields with increasing temperature 
\cite{Takeuchi_JPSJ70, Knafo_PRB81, Aoki_CRP14}.
This temperature dependence of PPM can be comparable to our results of $B_\mathrm{c}$.
For further understanding of the origin of $B_\mathrm{c}$, we need to measure high field and high temperature regions by various methods.

%%%% AdHvA

\subsection{
\label{subsect_AdHvA}
Acoustic de Haas-van Alphen effect
}

\begin{figure}
\begin{center}
\includegraphics[clip, width=0.50\textwidth]{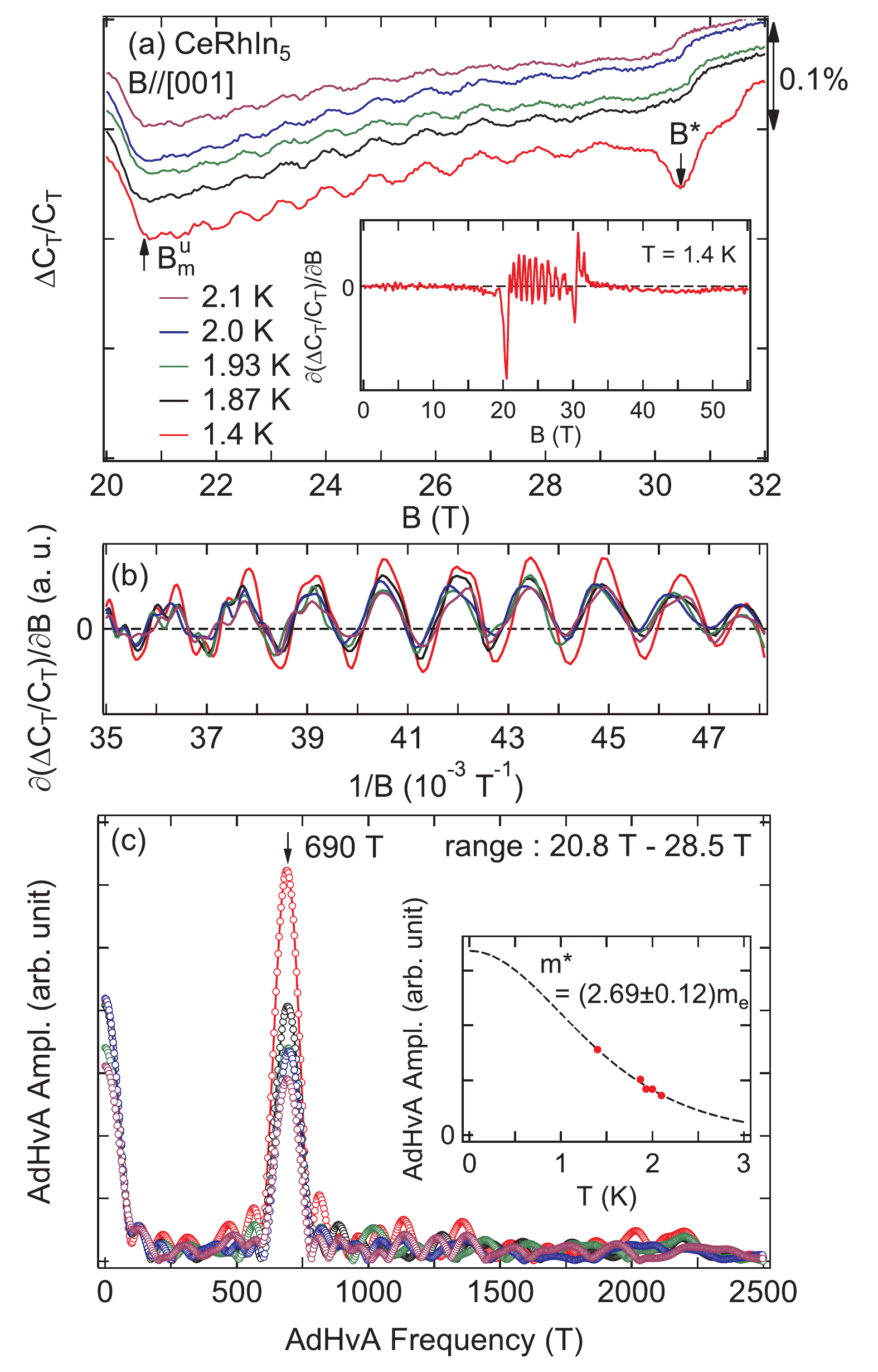}
\end{center}
\caption{
(Color online)
Acoustic de Haas-van Alphen oscillation observed in the transverse elastic constant $C_\mathrm{T} = \left( C_{11} - C_{12} \right)/2$ in CeRhIn$_5$ for $B // [001]$.
(a) Field dependence of $\mathit{\Delta}C_\mathrm{T}/C_\mathrm{T}$ in the field range between 20 T and 32 T at 1.4 K, 1.87 K, 1.93 K, 2.0 K and 2.1 K. 
The inset in panel (a) shows the first derivative of the relative elastic constant $\mathit{\Delta}C_\mathrm{T}/C_\mathrm{T}$ with respect to $B$ at 1.4 K.
(b) Inverse field dependence of $\partial \left( \mathit{\Delta}C_\mathrm{T}/C_\mathrm{T} \right)/\partial B$ at 1.4 K, 1.87 K, 1.93 K, 2.0 K and 2.1 K.
(c) Frequency dependence of the FFT amplitude of the AdHvA oscillation at different temperatures.
The inset in panel (c) shows the temperature dependence of AdHvA amplitude of 690 T.
The dashed line indicates the fit by the Lifshitz-Kosevich formula.  
}
\label{Fig4}
\end{figure}

To confirm our identification of the symmetry breaking, we focused on fermiology in terms of acoustic de Haas-van Alphen (AdHvA) effect by the transverse ultrasonic waves for $C_\mathrm{T}$. 
Figure \ref{Fig4}(a) shows the magnetic field dependence of the relative elastic constant $\mathit{\Delta}C_\mathrm{T}/C_\mathrm{T}$ at 1.4 K, 1.87 K, 1.93 K, 2.0 K and 2.1 K on field upsweep. 
We observed a clear AdHvA effect between $ B_\mathrm{m}^\mathrm{u} = 20.8$ T and $B^\star = 30$ T as indicated in the inset of Fig. \ref{Fig4}(a).
Here, a field tilting angle $\theta$ is estimated to be $5.5^\circ$ in this AdHvA measurements.
The first derivative of the relative elastic constant $\mathit{\Delta}C_\mathrm{T}/C_\mathrm{T}$ with respect to $B$ shown in Fig. \ref{Fig4}(b) exhibits a $1/B$ periodic behavior.
The appearance and the vanishing of AdHvA oscillation suggest that the FS reconstructions occur at the metamagnetic transition point $B_\mathrm{m}$ and at the EN transition point $B^\star$.

For a better understanding, we calculated the fast Fourier transformation (FFT) of the AdHvA oscillations.  
Figure \ref{Fig4}(c) shows the resultant FFT spectra obtained between 20.8 T and 28.5 T at several temperatures.
A sharp peak at 690 T and temperature dependence are shown in the FFT amplitude.
We can estimate the cyclotron mass $m^\star$ as $(2.69 \pm 0.12)m_e$ with the free-electron mass $m_e$ in terms of the Lifshitz-Kosevich formula
$A \left( T \right)
= A_0 \left( am^\star T/B \right) / \sinh \left( am^\star T/B \right) $,
which describes the temperature dependence of the amplitude of a quantum oscillation.
Here, $A_0$ is a constant, and $a$ can be written as $a = 2 \pi^2 k_\mathrm{B}/(e \hbar)$ using  the Boltzmann constant $k_\mathrm{B}$, the elementary charge $e$, and the Dirac constant $\hbar$.

%%%%%%%%%%%%%%%%%%%%%%%%Fig4%%%%%%%%%
%%%%%%%%%%%%%%%%%%%%%%%%%%%%%%%

Our ultrasonic measurements also indicate the contribution of the electric quadrupole to the AdHvA oscillation.
The amplitude of the AdHvA oscillation is also proportional to 
$\partial \left( \ln S_\mathrm{F} \right) / \partial \varepsilon_{ij}$
due to the deformation Hamiltonian given by 
\cite{Kataoka_JPSJ62, Settai_JPSJ63}
\begin{eqnarray}
\label{Hdef}
H_\mathrm{def} = \frac{1}{2}
\sum_{ij} \frac{ g_{ij} }{ \sqrt{m_i m_j} } p_i p_j \varepsilon_{ij}
,
\end{eqnarray}
where $S_\mathrm{F}$ is a cross-sectional area of the extremal FS, $g_{ij}$ for $i, j = x, y, z$ is the deformation coupling constant, $m_i(j)$ is the effective mass, and $\boldsymbol{p} = \hbar \boldsymbol{k}$ is the momentum of an electron around the Fermi level.
The Hamiltonian $H_\mathrm{def}$ in Eq. (\ref{Hdef}) is caused by the variation of the FS due to the strain $\varepsilon_{ij}$, and it can probably be attributed to the quadrupole-strain interaction in the $k$-space as discussed in Sect. \ref{sect_Quad}.
Thus, our experiment of AdHvA oscillation indicates that electrons on the FS with AdHvA frequency of 690 T and enhanced mass of $m^\star = (2.69 \pm 0.12)m_e$ have the electric quadrupole $O_{x^2-y^2}$, which induces the $B_\mathrm{1g}$ crystal symmetry breaking and the FS reconstruction due to the quadrupole-strain interaction.

This FS, however, has not been observed in previous dHvA measurements in magnetic fields along the [001] direction in CeRhIn$_5$
\cite{Shishido_JPSJ71, Moll_NatureComm7663, Jiao_PRB99, Cornelius_PRB62, Hall_PRB64, Elgazzar_PRB69}.
Nevertheless, the theoretical study treating $4f$ electrons as itinerant has shown a hole FS with frequency $\approx 690$ T, which is centered around the $\Gamma$ and $\mathrm{X}$ points of the Brillouin zone and constructed by the doubly-degenerate bands 90
\cite{Hall_PRB64}.
A similar hole FS existing around the $\Gamma$ point has been proposed by the theoretical calculations as the band-13 $\varepsilon$-branch in CeCoIn$_5$
\cite{Settai_JPhys13}
and 
the band-13 $g$-branch in CeIrIn$_5$
\cite{Elgazzar_PRB69, Haga_PRB63}.

These theoretical calculations on itinerant $4f$ indicate two things about the ultrasonic results in CeRhIn$_5$.
First, the FS with 690 T observed by the AdHvA oscillation can be measured without field tilting from $[001]$ direction.
In other words, the metamagnetic transition  would not change this FS.
Thus, the FS reconstruction can be induced at the EN phase.
Second, the field-induced itinerant character of the $4f$ electrons contributes to the FS.
As discussed in Sect. \ref{sect_Quad}, itinerant $4f$ character is important to consider the quadrupole effects.

To fully understand the shape of the FS and that of reconstruction, field angular dependence of AdHvA measurements are required.

%%%%%%%%%%%%%%%%   Discussion   %%%%%%%%%%%%%%%%%
\section{
\label{sect_Quad}
Quadrupole degree of freedom
}

In this section, we discuss the origin of the electric quadrupole to describe the field-induced EN phase.

First, we consider the quadrupole contribution in a zero field and high fields from the localized electron point of view.
In localized 4$f$-electron systems and related compounds, the electronic states under the CEF have been studied to describe the origin of an electric quadrupole and the elastic properties \cite{Tamaki_JMMM52, Nakamura_JPSJ63, Goto_JPSJ78, Baba_JPSJ80}.
It has been suggested that in CeRhIn$_5$ the CEF ground state can be described by the $\Gamma_7$ doublet, the first excited state is another $\Gamma_7$ doublet, and the second excited state is the $\Gamma_6$ doublet
\cite{Takeuchi_JPSJ70, Rosa_PRL122, Willers_PRB81}.
These CEF wavefunctions have expectation values of the electric quadrupole (see Appendix \ref{Appendix_A}).
Because the quadrupoles do not break time reversal symmetry, the degeneracy of each CEF-state, which is described as a Kramers doublet, is conserved.
Therefore, the Curie term in the quadrupole susceptibility caused by the diagonal elements of the $O_{x^2-y^2}$ matrix does not contribute to the elastic constant $C_\mathrm{T}$.
In addition to the Curie term, the van-Vleck term due to the off-diagonal elements in $\boldsymbol{O}_{x^2-y^2}$ in Eq. (\ref{Matrix of Ox2-y2}) also contributes to the quadrupole susceptibility.
However, the energy gap over 250 K
\cite{Takeuchi_JPSJ70, Willers_PRB81}
between the ground and the second excited states would be too wide to show the van-Vleck contribution in low temperatures where the EN phase appears.
These are the reasons why the EN transition does not appear in a zero field.
Since the Zeeman effect results in a mixing of the CEF states, the Curie and van-Vleck contributions can be enhanced by the magnetic fields.
However, the energy gap of 250 K would also be too large with respect to the energy scheme of the EN transition field $B^\star = 28.5$ T.

Therefore, we focus on the itinerant $4f$-electron character and the $p$-$f$ hybridization for an alternative explanation of the quadrupole degree of freedom.
The field-induced mass enhancement
\cite{Jiao_PRB99}
and the $p$-$f$ hybridization 
\cite{Rosa_PRL122}
indicated by the magnetostriction can be consistent with $4f$ delocalization. 
In addition to CeRhIn$_5$, the $B_\mathrm{1g}$-type crystal symmetry breaking and the quadrupole ordering have been revealed by the ultrasonic measurements in the iron pnictide compounds 
\cite{Goto_JPSJ80, Yoshizawa_JPSJ81, Kurihara_JPSJ86},
which is described appropriately as an itinerant-electron system.
Furthermore, the $B_\mathrm{1g}$-type lattice-instability driven by the $c$-$f$ hybridization has also been discussed for URu$_2$Si$_2$ 
\cite{Yanagisawa_PRB97}.
These ultrasonic results also indicate the importance of the quadrupole effects based on itinerant treatment and the $p$-$f$ hybridization of CeRhIn$_5$ in high fields.

For the itinerant $4f$-electron model, the electric quadrupole $O_{x^2-y^2}$ in $k$-space can be written using creation operators 
$\boldsymbol{c}_{ \boldsymbol{k}, l}^\dagger =  \left(c_{ \boldsymbol{k}, l}^\dagger, \cdots, c_{ \boldsymbol{k}, l'}^\dagger \right)$,  annihilation operators
$\boldsymbol{c}_{ \boldsymbol{k}, l} = \left(c_{ \boldsymbol{k}, l}, \cdots, c_{ \boldsymbol{k}, l'} \right)^\mathrm{T}$,
and the quadrupole matrix $\boldsymbol{O}_{x^2-y^2}$ in Eq. (\ref{Matrix of Ox2-y2}) in Appendix \ref{Appendix_A} as
%Quadrupole 2nd quant
\begin{align}
\label{Ox2-y2 2nd quant}
O_{x^2-y^2, \boldsymbol{k}, \boldsymbol{q} }
= \boldsymbol{c}_{ \boldsymbol{k} + \boldsymbol{q} , l}^\dagger \boldsymbol{O}_{x^2-y^2} \boldsymbol{c}_{ \boldsymbol{k}, l'} 
.
\end{align}
Here, $\boldsymbol{k}$ is the wave vector of the electron, $\boldsymbol{q}$ is the scattering vector, and $l$ and $l'$ are indices of the orbital of the electron.
For the quadrupole-strain interaction given in Eq. (\ref{HQS}), the scattering vector $\boldsymbol{q}$ in Eq. (\ref{Ox2-y2 2nd quant}) coincides with a wave number of phonons excited by ultrasound or heat.
A finite value of the electric quadrupole in Eq. (\ref{Ox2-y2 2nd quant}) probably results in the reconstruction of the FS and the crystal symmetry breaking at $B^\star$ due to the quadrupole-strain interaction in $k$-space, as given in Ref. \cite{Kurihara_JPSJ86}.
The band calculations for CeRhIn$_5$, CeIrIn$_5$, and CeCoIn$_5$ have shown the contribution of the itinerant $4f$ electrons to the energy band around the Fermi energy
\cite{Elgazzar_PRB69, Haga_PRB63, Rusz_PRB77}.
Thus, the itinerant behavior of CeRhIn$_5$ probably induces a finite value of the electric quadrupole in Eq. (\ref{Ox2-y2 2nd quant}).

In addition to the electric quadrupole formula in Eq. (\ref{Ox2-y2 2nd quant}), another description of $O_{x^2-y^2}$ in $k$-space is
$k_x^2 - k_y^2$
\cite{Hayami_PRB98, Watanabe_PRB98}.
This quadrupole formula probably enters in the deformation Hamiltonian in Eq. (\ref{Hdef}) for the strain $\varepsilon_{x^2-y^2}$ through the momentum $\boldsymbol{p} = \hbar \boldsymbol{k}$.
This is the reason for the AdHvA oscillation behavior of transverse elastic constant $C_\mathrm{T}$ in addition to the response at the EN phase, which indicates the quadrupole ordering of $O_{x^2-y^2}$.

Nevertheless, as indicated in the cases of CeIrIn$_5$ and CeCoIn$_5$, the itinerancy of $4f$ electrons can be attributed to the enhancement of the out-of-plane anisotropy of the CEF wavefunction
\cite{Willers_PNAS112, Chen_PRB97},
which is expected to induce an anomaly in the elastic constants $C_{33}$ and $C_{44}$.
Thus, for $B_\mathrm{1g}$-type in-plane anisotropy, not only the delocalization of $4f$ electrons need to be considered, but the in-plane $p$-$f$ hybridization as well
\cite{Rosa_PRL122}.
The expectation value of the electric quadrupole $O_\Gamma$, described as
$\int d\boldsymbol{r} \psi_l^\ast O_\Gamma \psi_{l'}$,
takes a non-zero value if both the $\psi_l$ and $\psi_{l'}$ wavefunctons have same parity for the coordinates $x$ and $y$
\cite{Hayami_PRB98}.
This symmetry consideration suggests that wavefunctions constructed by the Ce-$4f$ electrons and In-$5p$ electrons have a quadrupole degree of freedom.
In this treatment, $H_\mathrm{QS}$ in Eq. (\ref{HQS}) can be expanded by the quadrupole matrices 
$\boldsymbol{O}_{x^2-y^2}$ based on the $5p$ and $4f$ wavefunctions
and the creation and annihilation operators
$\boldsymbol{c}_{ \boldsymbol{k}, l}^\dagger$
and
$\boldsymbol{c}_{ \boldsymbol{k}, l}$ for the $l = p$ and $f$ orbitals.
It is expected that the field-induced in-plane $p$-$f$ hybridization enhances the quadrupole contribution to the susceptibility, the order parameter for the $B_\mathrm{1g}$-type EN transition and the crystal symmetry breaking, and the AdHvA oscillation in CeRhIn$_5$. 

To determine the origin of the field-induced EN transition in CeRhIn$_5$, theoretical studies for the FS, concerning the $p$-$f$ hybridization in high fields and the susceptibility of the electric quadrupole $O_{x^2-y^2, \boldsymbol{k}, \boldsymbol{q} }$, are required.
By determining the quadrupole-strain coupling constant, the phonon contribution to the EN transition can be better understood.

%%%%%%%%%%%%%%%%%%conclusion%%%%%%%%%%%%%%%

\section{
\label{conclusion}
Conclusion}

In the present work, we investigated the origin of the EN phase in high magnetic fields of CeRhIn$_5$ by the ultrasonic measurements.
We found a clear anomaly in the transverse elastic constant $C_\mathrm{T} = (C_{11} - C_{12})/2$ and a sharp peak in the ultrasonic attenuation coefficient $\alpha_\mathrm{T}$ at $B^\star$, while the anomaly in other elastic constants $C_{33}$, $C_{11}$, $C_{44}$ and $C_{66}$ were hardly visible.
This exhaustive measurement of the elastic constants indicates that the EN transition can be attributed to a $B_\mathrm{1g}$-type quadrupole ordering of $O_{x^2-y^2}$ with the $B_\mathrm{1g}$-type crystal symmetry breaking given by the strain $\varepsilon_{x^2-y^2}$ due to the quadrupole-strain interaction
$H_\mathrm{QS} =  -g_{x^2-y^2}O_{x^2-y^2}\varepsilon_{x^2-y^2}$.
A maximal non-isomorphic orthorhombic subgroup $Pmmm$ ($D_{2h}^1$) is an appropriate space-group for this symmetry lowering from $P4/mmm$ ($D_{4h}^1$) 
\cite{InternationalTables}.   
The AdHvA oscillation in $C_\mathrm{T}$ indicates a FS reconstruction accompanied by the EN transition in terms of vanishing AdHvA oscillation at $B^\star$.
The FS can be qualitatively explained by itinerant $4f$ electrons and the in-plane $p$-$f$ hybridization, which results from the electric quadrupole $O_{x^2-y^2}$ in $k$-space.
This can be the origin of the symmetry breaking in high fields and the AdHvA oscillation in CeRhIn$_5$.

\section*{Acknowledgment}
The authors thank Yuichi Nemoto and Mitsuhiro Akatsu for supplying the LiNbO$_3$ piezoelectric transducers and Keisuke Mitsumoto for valuable discussions.

 %%%%%%%%%%%%%%%%%%%%%Figure%%%%%%%%%%%%%%%%%%%%%%%%%%%%%%%

 %%%%%%%%%%%%%%%%%%Appendix%%%%%%%%%%%%%%%
\appendix

\section{
\label{Appendix_A}
Quadrupole matrix based on the CEF wave-functions
}

In this section, the wavefunction of CEF states and the quadrupole matrices used in Sect. \ref{sect_Quad} are presented. 
The wavefunctions of the CEF state can be written as \cite{Takeuchi_JPSJ70, Rosa_PRL122, Willers_PRB81}
%CEF wavefunctions
\begin{align}
\label{CEF WF_1}
\left| \Gamma_7^{\mathrm{G} \pm} \right>
&= \alpha \left| \pm \frac{5}{2} \right> + \sqrt{1-\alpha^2} \left|  \mp \frac{3}{2} \right>,
\\
\label{CEF WF_2}
\left| \Gamma_7^{1 \pm} \right>
&= \sqrt{1-\alpha^2} \left| \pm \frac{5}{2} \right> -\alpha \left|  \mp \frac{3}{2} \right>,
\\
\label{CEF WF_3}
\left| \Gamma_6^\pm \right>
&= \left| \pm \frac{1}{2} \right>,
\end{align}
where
$\left| \Gamma_7^{\mathrm{G} \pm} \right>$
are the ground states,
$\left| \Gamma_7^{1 \pm} \right>$
are the first excited states, and
$\left| \Gamma_6^\pm \right>$
are the second excited states.
The $\alpha$ value in Eqs. (\ref{CEF WF_1})-(\ref{CEF WF_3}) can be determined by the CEF parameters $B_2^0$, $B_4^0$, and $B_4^4$ in the CEF Hamiltonian under the tetragonal symmetry of $D_{4h}$.
Using the CEF wavefunctions and Stevens equivalent operator
$O_{x^2-y^2} = J_x^2 - J_y^2 = \left( J_+^2 + J_-^2 \right)/2$,
the matrix of the electric quadrupole $O_{x^2-y^2}$ in a zero field can be calculated as
%Matrix of Ox2-y2
\begin{align}
\label{Matrix of Ox2-y2}
\boldsymbol{O}_{x^2-y^2} 
= \bordermatrix{
& \Gamma_7^{\mathrm{G}+} & \Gamma_7^{\mathrm{G}-} & \Gamma_7^{1+} & \Gamma_7^{1-} & \Gamma_6^+ & \Gamma_6^- \cr
& 0 & 0 & 0 & 0 & \alpha_+ & 0 \cr
& 0 & 0 & 0 & 0 & 0 & \alpha_+ \cr
& 0 & 0 & 0 & 0 & \beta_- & 0 \cr
& 0 & 0 & 0 & 0 & 0 & \beta_-  \cr
& \alpha_+ & 0 & \beta_- & 0 & 0 & 0 \cr
& 0 & \alpha_+ & 0 & \beta_- & 0 & 0 
}.
\end{align}
Here, for the convenience, matrix elements in Eq. (\ref{Matrix of Ox2-y2}) are set as
$\alpha_\pm = \sqrt{10} \alpha \pm 3\sqrt{2} \sqrt{1-\alpha^2}$
and
$\beta_\pm = \sqrt{10} \sqrt{1-\alpha^2} \pm 3\sqrt{2} \alpha$.

%%%%%%%%%%%%%%%%%%%%%%%Ref%%%%%%%%%%%%%%%%%%%%%%%%%%%%%%%%

\end{document}